\documentclass[twocolumn,showpacs,preprintnumbers,amsmath,amssymb,aps,prl]{revtex4}
\usepackage{graphicx}
\begin{document}

\title{Shear Banding and Spatiotemporal Oscillations in Vortex Matter in 
Nanostructured Superconductors}   
\author{C. Reichhardt and    
 C. J. Olson Reichhardt} 
\affiliation{
 Theoretical Division,
Los Alamos National Laboratory, Los Alamos, New Mexico 87545} 

\date{\today}
\begin{abstract}
We propose a simple nanostructured pinning array geometry where
a rich variety of complex vortex shear banding phenomena 
can be realized.
A single row 
of pinning sites is removed
from a square pinning array. 
Shear
banding effects arise when vortex motion in the pin-free channel nucleates 
motion of vortices in the surrounding pinned regions, 
creating
discrete steps in the vortex velocity profile away from the channel. 
Near the global depinning transition, 
the width of the band of moving vortices 
undergoes
oscillations or fluctuations that can span the
entire system. 
We use simulations to show that these effects 
should be observable in the transport properties of the system. 
Similar large oscillations and shear banding effects are 
known to occur for sheared
complex fluids in which different dynamical phases coexist.     
\end{abstract}
\maketitle

\vskip2pc
Sheared complex fluids
\cite{Hoffman,Rouz,Britton,Berret,Fielding,Olmsted} and 
colloidal assemblies \cite{Coussot} exhibit the
dynamical phenomenon of {\it shear banding}, 
in which the motion of the system under an applied shear becomes 
stratified or concentrated in certain regions. For many complex fluids
where different dynamical flow phases can coexist, coupling of
these phases under shear  
can induce additional dynamics 
such as large oscillations and 
strongly fluctuating chaotic phases \cite{Fielding,Olmsted}. 
It would be interesting to explore whether similar behaviors can occur in 
other systems such as magnetic vortices in type-II superconductors.  
Shear banding induced by the presence of periodic structures, such as a
periodic pinning array for superconducting vortices, has not been
considered previously.
Shearing in complex fluids or colloids in the presence
of quenched disorder could be realized by having the flow traverse
an optical trap array or 
a periodic array of micron-scale obstacles.

In order to shear a superconducting vortex system, unconventional contact
geometries such as the Corbino geometry are normally used
\cite{Lopez,Zeldov,Marchetti,Okuma,Akria}. 
Simulations of pin-free systems 
using such geometries show 
transitions from rigid body rotation to liquid-like flow states 
\cite{Carmen}, and 
experiments 
revealed a similar effect near the vortex liquid 
to vortex solid transition \cite{Lopez}.
In other experiments, 
noise  and transport measurements of moving
vortices were performed in a Corbino geometry \cite{Zeldov,Okuma}. 
Recent simulations of Corbino contacts applied to very small superconducting
disks 
produced a rich
variety of novel vortex dynamics 
\cite{Misko}. 
In this work 
we propose the use of a simple periodic pinning array geometry rather
than a Corbino geometry to shear the vortices.
We observe a rich variety of shear banding effects including
an ordered phase with velocity steps,  giant spatiotemporal
oscillations, and strongly fluctuating chaotic phases. 

Using nanotechnology it is now possible to create tailored
pinning array structures 
in which the lattice symmetry and the order and size of the pinning sites can be
carefully controlled \cite{Baret,Martin}. 
It has also been demonstrated experimentally that these 
arrays can be
diluted by systematically removing individual pinning sites \cite{Kemmler}.
Here we propose removing entire rows from a square pinning array and applying a
magnetic field such that the vortex density matches 
the density of the original undiluted square pinning array.
Experimentally, this geometry could also be achieved on a larger scale by
periodically removing rows of pinning sites.  
We consider samples in which the pinning sites are sufficiently small that each
pin can capture only a single vortex; thus, 
when there are more vortices than pinning sites,
a portion of the vortices are located outside the pinning sites in
the pin-free regions.
Although this geometry is relatively simple, 
to our knowledge it has not been considered 
in previous simulations or experiments.  
Simulations of vortex dynamics in square pinning arrays 
near and just above the first matching field 
showed that the interstitial vortices are 
more mobile than the pinned vortices and depin at lower driving currents, while
coupling between the moving interstitial vortices and 
the pinned vortices can produce a series of 
transitions between different dynamical phases 
including a disordered or turbulent phase 
with a crossover to a laminar phase at higher
drives \cite{Reichhardt}. 
Although the simulations 
provided clear predictions for features in the transport curves 
associated with these transitions,
the experimental observation of these phases 
was not achieved until very recently
due to new experimental capabilities 
for reducing local heating effects \cite{Periodic} and for performing  
transport measurements at temperatures well below $T_{c}$ \cite{Periodic2}.
Molecular dynamics-style simulations should 
serve as excellent models for the experiments that are currently possible.

In this work we demonstrate 
that when a row or rows of pinning sites are removed
from a square pinning array, the vortex motion 
under an applied driving current first
occurs in the pin-free channel and nucleates the depinning and 
motion of the vortices in the pinning sites.
This produces a series of steps in the velocity response along with
fluctuating or chaotic phases and
giant temporal oscillations when a 
band of moving vortices propagates away from and toward the pin-free
channel.

We simulate a two-dimensional system of superconducting vortices
interacting with a modified square pinning array.
The sample is of size $L \times L$ and has
periodic boundary conditions in the $x$ and $y$ directions. 
The undiluted square pinning array contains $N_p$ pinning sites 
at a density of $n_p=N_p/L^2$, and the sample contains $N_v$ vortices at
a density of
$n_{v} = N_{v}/L^2$.
The field at which $N_v=N_p$ is the matching field
$B_{\phi}$.   
The dynamics of a vortex $i$ at position ${\bf R}_i$ 
is given by the overdamped equation of motion 
\begin{equation} 
\eta\frac{d{\bf R}_i}{dt} = {\bf F}^{vv}_{i} + {\bf F}^{p}_{i} + {\bf F}^{ext} .
\end{equation} 
The repulsive vortex-vortex interaction force is
${\bf F}^{vv}_{i} = 
\sum^{N_{v}}_{i\neq j}F_{0}K_{1}(R_{ij}/\lambda){\hat {\bf R}}_{ij}$.
Here $K_{1}$ is a modified Bessel function
that decays exponentially at large distances, 
$F_{0} = \phi_{0}/2\pi\mu_{0}\lambda^{3}$,     
$\lambda$ is the London penetration depth, $\phi_0=h/2e$ is the flux quantum, 
$\mu_{0}$ is the permeability of free space,
$R_{ij}=|{\bf R}_i-{\bf R}_j|$,
and ${\bf \hat R}_{ij}=({\bf R}_i-{\bf R}_j)/R_{ij}$. 
The pinning sites are modeled as parabolic traps with 
a maximum force of $F_{p}$ and radius $R_{p}=0.35\lambda$ 
which are placed in a square 
array with lattice constant $a$:
${\bf F}_i^p=\sum_k^{N_p}F_pF_0(R_{ik}/R_p)\Theta(R_p-R_{ik}){\bf \hat R}_{ik}$,
where $\Theta$ is the Heaviside step function.  
We set $n_p=0.28/\lambda^2$.
We keep $R_p$ and $F_p$ small enough that a maximum of one vortex can be 
captured by a given pinning site, and we remove one row of pins.
The initial vortex positions are obtained through simulated annealing, which
produces a square vortex arrangement at the matching field $B=B_\phi$.
The external driving force 
${\bf F}^{ext} = F_{D}{\hat {\bf x}}$
represents the Lorentz force from an applied current. 
We increase $F_D$ in small increments
and measure the vortex velocity averaged over 
$\tau=8000$ simulation time steps,
$\langle V_x\rangle =N_v^{-1}\tau^{-1}\sum_t^{\tau}\sum_i^{N_v}d{\bf R}_i(t)/dt \cdot {\bf \hat x}$.

\begin{figure}
\includegraphics[width=3.5in]{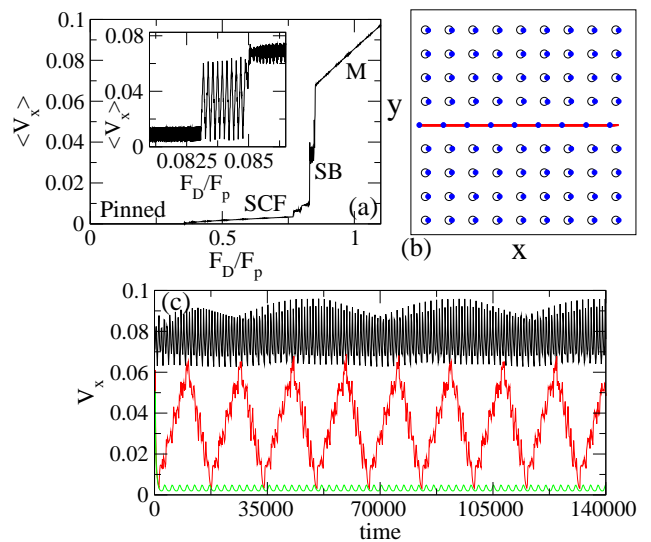}
\caption{
(a) The average velocity $\langle V_x\rangle$ 
vs applied force $F_{D}/F_{p}$ for 
a system with 
$F_{p} = 0.1$, 
$B/B_{\phi} = 1.0$, and a single removed row of pinning. 
Four regimes are marked: a pinned regime, a single channel flow
(SCF) regime, a shear banding (SB) regime,
and a moving (M) phase in which all of the vortices are flowing. 
Inset: A blowup of the main panel 
for $0.081 \le F_D/F_p \le 0.0865$ in the shear banding phase 
where a short time average is used for the velocities 
so that the fluctuations can be seen more clearly. 
Large scale oscillations occur as the system switches back and forth
between three moving rows of vortices and all rows of vortices moving.
On a long time average, the oscillating regime has an intermediate 
value of $\langle V_x\rangle$ as shown in the main panel. 
(b) The vortex positions (dots), trajectories (lines), and
pinning site locations (open circles)
for a small subsection of the sample in (a) 
in the SCF phase at $F_{D}/F_{p} = 0.6$.
(c) $V_x$ vs time for the system in (a) 
at $F_{D}/F_{p} = 0.7$, 0.85, and $0.92$ (from bottom to top).          
}
\end{figure}

In Fig.~1(a) we plot the velocity $\langle V_x\rangle$ versus 
force $F_{D}/F_p$ curve for a sample with 
$F_{p} = 0.1$
and $B/B_{\phi} = 1.0$.
The central portion of the sample is illustrated in Fig.~1(b)
at
$F_D/F_p=0.6$, where only the interstitial vortices in the pin-free row
are in motion. 
There are four distinct regimes in Fig.~1(a). In the pinned  
regime at low $F_{D}$,
the interstitial vortices in the pin-free channel remain pinned due to 
the repulsive interactions from the vortices at the pinning sites. 
At higher drives, the interstitial vortices depin and 
enter the single channel flow (SCF) state
illustrated in Fig.~1(b). 
A sharp increase in $\langle V_{x}\rangle$ occurs near $F_{D}/F_{p} = 0.75$
and corresponds to the onset of motion of the vortices 
in pinning rows adjacent to the pin-free channel,
termed the shear banding (SB) regime. 
The velocity of the vortices moving 
along the pinning rows is less than the velocity of the vortices
in the pin-free channel, and during the time interval required for 
the vortices in the pinning rows to move a distance $a$ and
hop from one pinning site to another, 
the vortices in the pin-free channel move a distance $2a$ or greater.
A small step in $\langle V_x\rangle$ appears
when the motion of the vortices in the adjacent 
pinning sites locks into step 
with the motion of the vortices in the pin-free channel. 
For $F_{D}/F_{p} > 0.855$ all the vortices are mobile and
$\langle V_x\rangle$ increases linearly with increasing $F_{D}/F_p$. 
For $0.83 \le F_D/F_p \le 0.855$, $\langle V_x\rangle$ takes a value that is
close to half of the Ohmic value,
indicating that on average half of the vortices in the sample 
are moving. 
It is not clear from the time-averaged data whether this means that half of 
the vortices are always moving and the other half are always pinned, or
whether the number of moving vortices is fluctuating.
To clarify this, in the inset of Fig.~1(a) we plot $\langle V_x\rangle$
over a small range of $F_D/F_p$ using a shorter time average of $\tau=175$
simulation time steps.
For $0.83 < F_{D}/F_p < 0.85$, this $\langle V_x\rangle$
response exhibits large oscillations 
with a maximum velocity close to the Ohmic value and a
minimum velocity close to that associated with vortices flowing in three
rows,
indicating that the number of
moving vortices is oscillating over time. 
In Fig.~1(c) we plot the instantaneous $V_x$ versus time
at fixed $F_{D}/F_{p} = 0.7$, 0.85, and $0.92$.
For $F_{D}/F_{p} = 0.7$, only three vortex rows 
are moving and there is a high frequency periodic component
of $V_x$ produced by the motion of the vortices over the
periodic pinning sites.
For $F_{D}/F_{p} = 0.92$, all of the vortices are moving and there is
again a high frequency periodic signal in $V_x$ caused by the
periodic motion over the pinning sites.
At $F_{D}/F_{p} = 0.85$, 
pronounced oscillations appear in $V_x$ that have
a much larger amplitude and a lower frequency, indicating that these
oscillations are not associated with the periodicity of the pinning 
sites. 
A high frequency oscillatory component of $V_x$ which is a result of the 
motion over the periodic pinning sites does appear; however,
the larger scale oscillations 
are due to the emission and absorption of a transverse front of moving vortices
originating at the pin-free channel.

\begin{figure}
\includegraphics[width=3.5in]{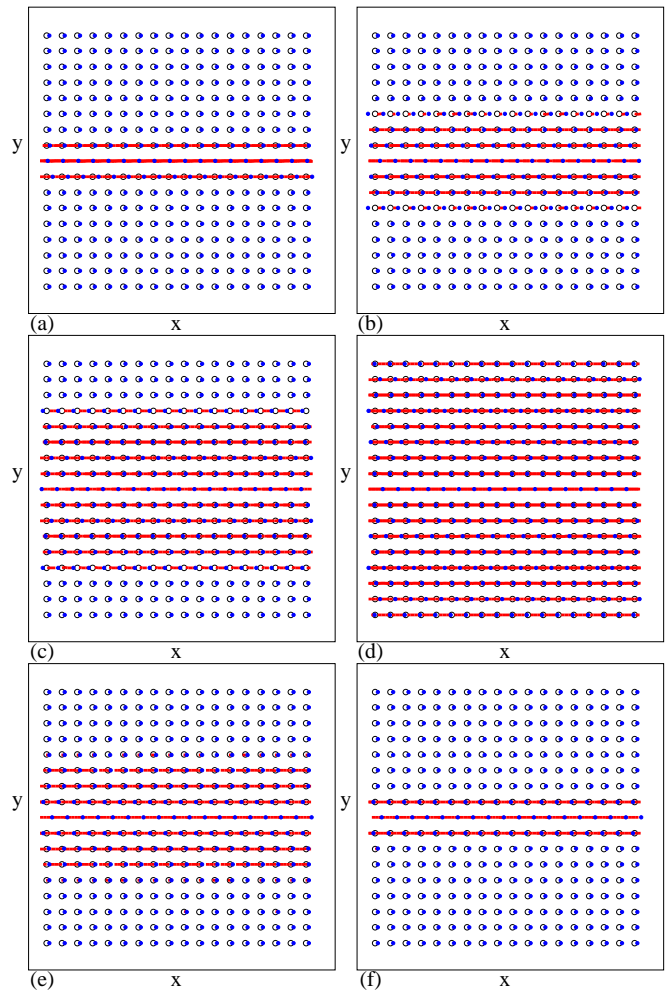}
\caption{
The vortex positions (dots), trajectories (lines), and pinning site locations 
(circles)
for the system in Fig.~1(c) at $F_{D}/F_{p} = 0.85$ during consecutive
time intervals in a single large oscillation of $V_x$ with period $T$.
(a) At a minimum in $V_x$, $t/T=0$, 
the vortices in the pin-free channel move along
with vortices in one pinned row on each side of the channel.
(b) At $t/T=0.125$, a front of moving rows begins to propagate outwards
from the pin-free channel and here
three pinned rows on each side of the channel are moving.
(c) Near $t/T=0.25$ the moving region continues 
to grow.  
(d) At $t/T=0.5$ the front has moved 
completely through the sample and all of the vortices are
moving.           
(e) Near $t/T=0.8$ the moving region is contracting.
(f) At $t/T=1$ the system returns to the minimum state of three moving 
rows of vortices. 
}
\end{figure}

In Fig.~2 we illustrate the vortex trajectories during consecutive time 
intervals for the system shown in
Fig.~1(c) at $F_{D}/F_{p} = 0.85$ 
for a single cycle of the large oscillations in $V_x$, which has a
period of $T=15750$ simulation time steps.
At the start of the cycle, $V_x$ is at its minimum value 
and the vortices in the
pin-free channel are moving along with vortices in the two adjacent pinning 
rows.  Figure~2(a-c) shows that during the first half of the oscillation,
$t/T<0.5$, an increasing number of vortices in the pinning rows become mobile, 
and there is a well defined front between the moving and non-moving 
portions of the sample.  At the halfway point of the oscillation,
$t/T=0.5$, Fig.~2(d) indicates that all of the vortices are mobile.
When the two depinning fronts meet, 
an instability is triggered that causes the vortices to 
become pinned again, and two pinning fronts begin to propagate back 
toward the pin-free channel during the second half of the period, $0.5<t/T<1$.
Finally at $t/T=1$, the system returns to the initial state 
containing only three moving rows of vortices. The period $T$ of the
oscillations at fixed $F_{D}$ 
depends on the system size since the depinning and 
pinning fronts propagate at a fixed speed.
By simulating large systems, we find that $T$ scales with $L$. 
The high-frequency periodic component of the vortex motion that is
generated when the vortices move over the periodic pinning array does
not change with system size and is a function only of the 
pinning lattice constant $a$.
Shear banding effects resulting in large oscillations have been observed
for shearing in many complex fluids and are due to 
a nonlinear coupling between two different types
of coexisting structures that form in the 
fluid \cite{Fielding,Olmsted}. In our vortex system the
two structures are the
square pinned vortex lattice and the distorted hexagonal ordering of the
moving vortices.  The mismatch between these two phases results in the
formation of topological defects at the boundary between the phases.
When the depinning fronts meet, 
it is possible that a portion of the defects annihilate, changing the
direction of propagation of the front. 

\begin{figure}
\includegraphics[width=3.5in]{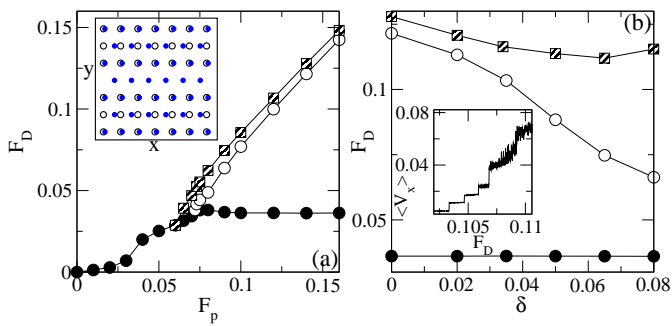}
\caption{ 
The dynamic phase diagram $F_{D}$ vs $F_{p}$ for the system in Fig.~1. 
Filled circles: depinning transition.  Open circles: transition
from the 
SCF to the 
SB state. 
Hatched squares: transition from the 
SB state to the 
moving state. 
Inset: Vortex positions (filled circles) and pin positions (open circles) in
a small portion of the sample illustrating the depinning
process for $F_{p} < 0.6$.  Depinning
occurs in a single step from a partially pinned elastic solid  
where half of the vortices are shifted off the pinning sites, as shown.
(b) The phase diagram $F_D$ vs $\delta$ for the same system 
with $F_{p} = 0.14$, where $\delta$ is a measure of
strength 
dispersion for the pinning sites. As $\delta$ increases, the width of the
SB region grows. 
Inset: For $\delta = 0.035$, there are more steps
in $\langle V_x\rangle$ vs $F_D$
at the onset of the shear banding and the large oscillations are replaced with
a fluctuating or chaotic regime.
}
\end{figure}
    
In Fig.~3(a) we show a phase diagram of $F_{D}$ versus $F_{p}$ highlighting 
the different phases. 
For weak pinning, $F_{p} < 0.06$, the depinning transition occurs elastically 
in a single step where all of the
vortices begin moving at the same time. 
The effect of removing one pinning row is negated by the fact that
the vortex ground state for weak pinning forms a
distorted triangular lattice by shifting half the vortices
off the pinning sites, as shown in the inset of Fig.~3(a). 
The presence of the pin-free channel merely removes
the degeneracy of the ground state 
and ensures that this shift occurs along the $x$-direction.
For $F_{p} > 0.078$, the initial depinning occurs plastically and the vortex
motion first occurs in the pin-free channel.
The depinning force saturates with increasing
$F_{p}$ since the vortices in the pin-free 
channel are pinned by their interactions
with the neighboring pinned vortices, and this interaction strength does not
depend on $F_p$.
The transition to the moving phase 
shifts linearly to higher $F_D$ with increasing $F_{p}$. 
Open circles in Fig.~3(a) indicate the transition from the
SCF to the SB giant oscillation regime where multiple rows of vortices
are moving.
For strong pinning, the SB oscillation region is narrow; however, 
the addition of disorder or an increase in the 
size of the pinning sites can extend 
the width of the SB region. 
To examine the effect of disorder,
we changed the pinning strength from uniform to a normal distribution
with mean $F_p$ and width $\delta$.
In Fig.~3(b), we plot the onset of the different phases for $F_{D}$ 
versus $\delta$
for a system with $F_{p} = 0.14$. As $\delta$ increases, the depinning
transition to the SCF regime is unchanged since the
depinning force is determined by the 
fixed vortex-vortex interaction strength. 
The onset of the 
SB regime
shifts to higher $F_D$ with increasing disorder.  
For $\delta < 0.035$, the dynamics 
are the same as those 
illustrated in Fig.~1(a); however, for larger $\delta$ the shear banding 
occurs in a series of steps and the large oscillation region is replaced with
a strongly fluctuating or chaotic regime. 
In the inset of Fig.~3(b) we plot $\langle V_x\rangle$ 
vs $F_{D}$ near the onset of the 
SB regime for $\delta = 0.035$.
Here there are four steps 
that correspond to the opening of additional mobile rows of vortices.
The steps are followed by a transition to a more strongly fluctuating
regime where additional moving rows open and close intermittently.

In summary, we propose a simple periodic pinning geometry 
for vortices in superconductors in which 
a rich variety of complex shear banding phenomena can be realized.
A single row of pinning sites is removed to create an easy flow
channel.  Under an applied drive the vortices in the channel 
move first but due to their coupling with the vortices in the adjacent
pinning sites, vortex motion along the neighboring rows is also
nucleated.
As the global depinning
transition is approached, bands of moving rows of vortices 
propagate away from and toward
the channel, creating giant spatio-temporal 
oscillations which can be seen in transport signatures.
As disorder is added to the periodic pinning array, 
the velocity-force curves develop discrete steps and
the oscillating phase is replaced by a strongly fluctuating phase. 
These effects are similar
to shear banding phenomena in complex fluids where different dynamical phases
can coexist and couple.

This work was carried out under the auspices of the 
NNSA of the 
U.S. DoE
at 
LANL
under Contract No.
DE-AC52-06NA25396.

\end{document}